\documentstyle[prl,floats,aps,twocolumn,epsf,graphicx]{revtex}
\begin{document}
\twocolumn[\hsize\textwidth\columnwidth\hsize\csname
@twocolumnfalse\endcsname

\title{Asymptotic quasinormal mode spectrum of rotating black holes}
\author{Shahar Hod}
\address{Department of Condensed Matter Physics, Weizmann Institute, Rehovot 76100, Israel}
\date{\today}
\maketitle

\begin{abstract}

\ \ \ Motivated by novel results in the theory of black-hole quantization, we study 
{\it analytically} the quasinormal modes (QNM) of ({\it rotating}) Kerr black holes. The 
black-hole oscillation frequencies tend to the asymptotic value $\omega_n=m\Omega+i2\pi T_{BH}n$ 
in the $n \to \infty$ limit. This simple formula is in agreement with 
Bohr's correspondence principle. Possible implications of this
result to the area spectrum of quantum black holes are discussed.
\end{abstract}
\bigskip

]

Gravitational waves emitted by a perturbed black hole are dominated by
`quasinormal ringing', damped oscillations with a {\it discrete}
spectrum (see e.g., \cite{Nollert1} for a detailed review). 
At late times, all perturbations are
radiated away in a manner reminiscent of the last pure dying tones of
a ringing bell \cite{Press,Cruz,Vish,Davis}. 

Being the characteristic `sound' of
the black hole itself, these free oscillations 
are of great importance from 
the astrophysical point of view. They allow a direct way of identifying the spacetime 
parameters (especially, the mass and angular momentum of the central black hole). 
This has motivated a flurry of activity with the aim of computing the spectrum of oscillations.

The ringing frequencies are located 
in the complex frequency plane characterized by Im$\omega >0$. It turns
out that for a given angular harmonic index $l$ there exist an 
infinite number of quasinormal modes, for $n=0,1,2,\dots$, characterizing oscillations with decreasing
relaxation times (increasing imaginary part) \cite{Leaver,Bach}. On 
the other hand, the real part of the frequencies approaches an asymptotic 
{\it constant} value.

The QNM have been the subject of much recent attention (see e.g., \cite{Dreyer,Kun,Motl,Cor,MotlNei,CarLem,Hod2,KaRa,Abd,Ber1,Cho,Bri1,Gla,Nei,Mol,Pol,Bri2,LiHui,CarKon,BirCar,Swa,Pad,Bir,Iiz,Ber2,AndHow} and 
references therein), with the hope that these classical frequencies
may shed some light on the elusive theory of quantum gravity. 
These recent studies are motivated by an earlier work of Hod \cite{Hod1}. 

Few years ago I proposed to apply 
{\it Bohr's correspondence principle} in order to determine the value of the 
fundamental area unit in a quantum theory of gravity. 
It is useful to recall that in the early development of quantum mechanics, Bohr 
suggested a correspondence between classical and quantum properties of the
Hydrogen atom, namely that ``transition frequencies at large quantum numbers should equal
classical oscillation frequencies''. The black hole is in many senses 
the ``Hydrogen atom'' of General relativity. I therefore suggested \cite{Hod1} a similar usage of the 
discrete set of black-hole frequencies in order to shed some light on the {\it quantum} 
properties of a black hole. There is, however one important difference between the Hydrogen atom and 
a black hole: while a (classical) atom emits radiation spontaneously according to the (classical) 
laws of electrodynamics, a {\it classical} black hole does not emit radiation. This crucial 
difference hints that one should look for the highly damped black-hole free oscillations 
[let $\omega=$Re$\omega+i$Im$\omega$, then 
$\tau \equiv ($Im$\omega)^{-1}$ is the effective 
relaxation time for the black hole to return to a quiescent
state after emitting gravitational radiation. Hence, the relaxation time $\tau \to 0$ 
as Im$\omega \to \infty$, implying no radiation emission, as should be the case for a 
classical black hole].

Leaver \cite{Leaver} was the first to address to problem of computing the black hole 
highly damped ringing frequencies. Nollert \cite{Nollert2} found numerically 
(see also \cite{Andersson}) that the asymptotic
behavior of the ringing frequencies of a Schwarzschild black hole is
given by (we normalize $G=c=2M=1$)

\begin{equation}\label{Eq1}
\omega_n=0.0874247+{i \over 2} \left(n+{1 \over 2} \right)\  ,
\end{equation}
In \cite{Hod1} it was realized that 
this asymptotic value equals $\ln3 /(4\pi)$. 
A heuristic picture (based on thermodynamic and statistical physics arguments) was suggested 
trying to explain this fact \cite{Hod1}. 
Most recently, Motl \cite{Motl} has given an analytical proof for this equality.
 
Using the relation $A=16 \pi M^2$ for the surface area of a Schwarzschild black hole, 
and $\Delta M=E=\hbar \omega$ one 
finds $\Delta A=4{\ell^2_P} \ln3$ with the emission of a quantum, where 
$\ell_P$ is the Planck length. 
Thus, we concluded that the area
spectrum of the quantum Schwarzschild black hole is given by
 
\begin{equation}\label{Eq2}
A_n=4 {\ell^2_P} \ln 3 \cdot n\ \ \ ;\ \ \ n=1,2,\ldots\ \  . 
\end{equation}

This result is remarkable from a statistical physics point of
view. In the spirit of Boltzmann-Einstein formula in
statistical physics, Mukhanov and Bekenstein \cite{Muk,BekMuk,Beken2,Beken3}
relate $g_n \equiv exp[S_{BH}(n)]$ to the number of microstates of the
black hole that correspond to a particular external macrostate. 
In other words, $g_n$ is the
degeneracy of the $n$th area eigenvalue. The accepted thermodynamic
relation between black-hole surface area and entropy $S_{BH}={1 \over 4}A$ \cite{Beken1}, 
combined with the requirement that $g_n$ has to be an integer for every $n$, actually 
enforce a factor of the form $4\ln k$ 
(with $k=2,3,\dots$) in Eq. (\ref{Eq2}). It turns out that the value $k=3$ is the only one 
compatible both with the area-entropy thermodynamic relation for black hole, and with 
Bohr's correspondence principle as well.

It should be emphasized that the asymptotic behavior of the black hole ringing
frequencies is known only for the simplest case of a Schwarzschild 
black hole. Less is known about the corresponding 
QNM spectrum of the (rotating) Kerr black hole, in which case numerical calculations are limited 
to $n \leq 50$ \cite{Leaver,Det,Ono,Ber1,Ber2}. This is a direct consequence of the numerical 
complexity of the problem. The aim of the present Letter is to study analytically the asymptotic 
quasinormal mode spectrum of generic Kerr black holes.

The black-hole perturbations are governed by the well-known 
Regge-Wheeler equation \cite{RegWheel} in the case of 
a Schwarzschild black hole, and by the Teukolsky equation \cite{Teukolsky} for the 
(rotating) Kerr black hole. The black hole QNM correspond to
solutions of the wave equations with the physical boundary
conditions of purely outgoing waves at spatial infinity 
and purely ingoing waves crossing the event horizon \cite{Detwe}. Such boundary 
conditions single out {\it discrete} solutions $\omega$ (assuming a time dependence of the 
form $e^{i\omega t}$). 
The solution to the radial Teukolsky equation may be expressed as \cite{Leaver}

\begin{eqnarray}\label{Eq3}
R_{lm}& = &e^{i\omega r} (r-r_-)^{-1-s+i\omega+i\sigma_+} (r-r_+)^{-s-i\sigma_+}\nonumber \\
&&\Sigma_{n=0}^{\infty} d_n \Big({{r-r_+} \over {r-r_-}}\Big)^n\  ,
\end{eqnarray}
where $r_{\pm} =M \pm (M^2-a^2)^{1/2}$ are the black hole (event and inner) horizons 
($a=J/M$ is the black hole angular momentum per unit mass). 
The field spin-weight parameter $s$ takes the values $0, -1, -2$ respectively, 
for scalar, electromagnetic and gravitational fields. 

The sequence of expansion coefficients $\{d_n:n=1,2,\ldots\}$ is determined by a 
recurrence relation:

\begin{equation}\label{Eq4}
\alpha_n d_{n+1}+\beta_n d_n +\gamma_n d_{n-1}=0\  ,
\end{equation}
where $d_0=1$ and
 
\begin{equation}\label{Eq5}
\alpha_0 d_1+\beta_0 d_0=0\  .
\end{equation}

The recursion coefficients are given by \cite{Leaver}

\begin{equation}\label{Eq6}
\alpha_n=n^2+(c_0+1)n+c_0\  ,
\end{equation}

\begin{equation}\label{Eq7}
\beta_n=-2n^2+(c_1+2)n+c_3\  ,
\end{equation} 
and 

\begin{equation}\label{Eq8}
\gamma_n=n^2+(c_2-3)n+c_4-c_2+2\  ,
\end{equation} 
where the intermediate constants $c_n$ are defined by

\begin{equation}\label{Eq9}
c_0=1-s-i\omega-{{2i} \over b}\Big({\omega \over 2} -am\Big)\  ,
\end{equation}

\begin{equation}\label{Eq10}
c_1=-4+2i\omega(2+b)+{{4i} \over b}\Big({\omega \over 2} -am\Big)\  ,
\end{equation}

\begin{equation}\label{Eq11}
c_2=s+3-3i\omega-{{2i} \over b}\Big({\omega \over 2} -am\Big)\  ,
\end{equation}

\begin{eqnarray}\label{Eq12}
c_3& = &\omega^2(4+2b-a^2)-2am\omega-s-1+(2+b)i\omega\nonumber \\
&&-A_{lm}+{{4\omega+2i} \over b}\Big({\omega \over 2} -am\Big)\  ,
\end{eqnarray}
and 

\begin{eqnarray}\label{Eq13}
c_4& =&s+1-2\omega^2-(2s+3)i\omega\nonumber \\
&&-{{4\omega+2i} \over b}\Big({\omega \over 2} -am\Big)\  ,
\end{eqnarray}
where $b=(1-4a^2)^{1 \over 2}$ and 
the separation constants $A_{lm}$ are given by an independent recurrence relation \cite{Leaver}.

The quasinormal frequencies are determined by the requirement that the 
series in Eq. (\ref{Eq3}) absolutely convergent, i.e. that is $\Sigma d_n$ exists and 
is finite \cite{Leaver}.

The {\it physical} content of the expansion coefficients becomes clear when they are 
expressed in terms of the black-hole physical parameters 
$T_{BH}$ and $\Omega$, in which case they obtain a surprisingly 
simple (and compact) form \cite{Note1}

\begin{equation}\label{Eq14}
\alpha_n=-i{{n+1} \over {2\pi T_{BH}}}\Big[\omega-m\Omega+i2\pi T_{BH}(n+1-s)\Big]\  ,
\end{equation}
and 

\begin{equation}\label{Eq15}
\gamma_n=-{{2\omega +in} \over {2\pi T_{BH}}}\Big[\omega-m\Omega+i2\pi T_{BH}(n+s)\Big]\  .
\end{equation} 
Here $T_{BH}=(r_{+}-r_{-})/A$ is the Bekenstein-Hawking temperature, 
and $\Omega=4 \pi a /A$ is the angular velocity of the black-hole horizon. 
The explicit expression of the $\{\beta_n\}$ coefficients is not important for the analysis, 
but for later purposes it is important to note that $\beta_n \to -(a\omega)^2$ in the 
$n \to \infty$ limit with $\omega \sim -i2\pi T_{BH}n$ \cite{Note2}.

{\it Quasinormal frequencies of the Kerr black hole.} Taking cognizance of Eqs. (\ref{Eq14}) 
and (\ref{Eq15}), one finds that 
in the $n \to \infty$ limit with $\omega \sim -i2\pi T_{BH}n$, the $\alpha_n$ and $\gamma_n$ coefficients 
are of order $O(n)$, while $\beta_n=O(n^2)$. The 
term $\alpha_n d_{n+1}$ in Eq. (\ref{Eq4}) is therefore negligible as compared to the 
other two terms $\beta_n d_n$ and $\gamma_n d_{n-1}$ [one finds 
$\alpha_n d_{n+1}/\beta_n d_n=O(a^{-4}b^{-4}n^{-2})$. The asymptotic limit therefore requires 
$n \gg (ab)^{-2}$]. 
If $\gamma_N=0$ for some 
integer $N$, then {\it all} $d_n$ with $n \geq N$ values would vanish, implying 
the convergence of the series $\sum d_n$. 
Taking cognizance of Eq. ({\ref{Eq15}), one finds that $\gamma_N=0$ for $\omega_N=m\Omega-i2\pi T_{BH}N$. 
Thus, the asymptotic quasinormal frequencies of a rotating Kerr black hole are 
given by the simple relation
 
\begin{equation}\label{Eq16}
\omega_n=m\Omega-i2\pi T_{BH}n\  .
\end{equation}

We note that this asymptotic formula is in agreement with recent numerical results \cite{Ber2} for 
the $l=m=2$ gravitational perturbations. 
However, the authors of \cite{Ber2} 
cannot exclude the possibility that their numerical calculations for other perturbations 
actually break down {\it before} reaching the 
asymptotic regime \cite{Ber2}.

Taking cognizance of the first law of black-hole thermodynamics 

\begin{equation}\label{Eq17}
\Delta M=T_{BH} \Delta S + \Omega \Delta J\  ,
\end{equation}
one finds that the asymptotic frequency corresponds to $\Delta S=\Delta A=0$ \cite{Ber2}. In other words, the 
application of Bohr's correspondence principle to black-hole physics is {\it consistent} with 
the first law of black-hole thermodynamics.

{\it Quasinormal frequencies of a Schwarzschild black hole.} An analysis along the 
same lines as before provides a simple and elegant way to obtain 
the asymptotic quasinormal spectrum of electromagnetic perturbations of the 
spherically symmetric Schwarzschild black hole. 
One finds that in this case $\alpha_n=O(n)$ in the $n \to \infty$ limit 
with $\omega \sim -in/2$, while the $\beta_n$ and $\gamma_n$ 
coefficients are of order $O(n)$. Thus, the term $\beta_n d_n$ in Eq. (\ref{Eq4}) is 
negligible as compared to the 
other two terms $\alpha_n d_{n+1}$ and $\gamma_n d_{n-1}$ [one finds 
$\beta_n d_n/\alpha_n d_{n+1}=O(n^{-{1 \over 2}})$]. 
If $\gamma_N=0$ for some even (odd) 
integer $N$, then {\it all} $d_n$ with even (odd) $n>N$ values would vanish. 
Taking cognizance of Eq. ({\ref{Eq15}), one finds that for 
$\omega_N=-iN/2$ and $s=-1$ (electromagnetic perturbations), 
{\it both} $\gamma_N$ and $\gamma_{N+1}$ vanish, 
implying $d_N=0$ for all $n>N$ values. This would guarantee the convergence of the series $\sum d_n$ (the 
same conclusion holds true for all odd integer values of the parameter $|s|$). 
The electromagnetic ($s=-1$) quasinormal frequencies of a Schwarzschild black hole are therefore 
given asymptotically by the simple relation
 
\begin{equation}\label{Eq18}
\omega_n=-in/2\  .
\end{equation}
which agrees with the analysis of \cite{Motl}.

In summary, we have studied analytically the quasinormal mode spectrum of 
rotating Kerr black holes. The oscillation frequencies tend to the asymptotic value 
$\omega=m\Omega+i2\pi T_{BH}n$ in the $n \to \infty$ limit. This formula is 
consistent with Bohr's correspondence principle, 
and gives further support to its applicability in the quantum theory of gravitation. 

\bigskip
\noindent
{\bf ACKNOWLEDGMENTS}
\bigskip

I thank a support by the 
Dr. Robert G. Picard fund in physics. 
This research was supported by grant 159/99-3 from the Israel Science Foundation.


\begin{thebibliography}{99}

\bibitem{Nollert1} For an excellent review and a detailed list of references see 
H. P. Nollert, Class. Quantum Grav. {\bf 16}, R159 (1999).

\bibitem{Press} W. H. Press, Astrophys. J. {\bf 170}, L105 (1971).

\bibitem{Cruz} V. de la Cruz, J. E. Chase and W. Israel,
  Phys. Rev. Lett. {\bf 24}, 423 (1970).

\bibitem{Vish} C.V. Vishveshwara, Nature {\bf 227}, 936 (1970).

\bibitem{Davis} M. Davis, R. Ruffini, W. H. Press and R. H. Price,
  Phys. Rev. Lett. {\bf 27}, 1466 (1971).

\bibitem{Leaver} E. W. Leaver, Proc. R. Soc. A {\bf 402}, 285 (1985).

\bibitem{Bach} A. Bachelot and A. Motet-Bachelot,
  Ann. Inst. H. Poincar\'e {\bf 59}, 3 (1993).

\bibitem{Dreyer} O. Dreyer, Phys. Rev. Lett. {\bf 90}, 081301 (2003).

\bibitem{Kun} G. Kunstatter, Phys. Rev. Lett. {\bf 90}, 161301 (2003).

\bibitem{Motl} L. Motl, e-print gr-qc/0212096.

\bibitem{Cor} A. Corichi, Phys. Rev. D {\bf 67}, 087502 (2003).

\bibitem{MotlNei} L. Motl and A. Neitzke, e-print gr-qc/0301173.

\bibitem{CarLem} V. Cardoso and  J. P. S. Lemos, Phys. Rev. D {\bf 67}, 084020 (2003).

\bibitem{Hod2} S. Hod, Phys. Rev. D {\bf 67}, 081501 (2003).

\bibitem{KaRa} R. K. Kaul and S. K. Rama, e-print gr-qc/0301128.

\bibitem{Abd} E. Abdalla, K. H. C. Castello-Branco, and A. Lima-Santos, 
e-print gr-qc/0301130.

\bibitem{Ber1} E. Berti and K. D. Kokkotas, e-print hep-th/0303029.

\bibitem{Cho} H. T. Cho, e-print gr-qc/0303078.

\bibitem{Bri1} A. M. van den Brink, e-print gr-qc/0303095.

\bibitem{Gla} K. Glampedakis and N. Andersson, e-print gr-qc/0304030.

\bibitem{Nei} A. Neitzke, e-print hep-th/0304080.

\bibitem{Mol} C. Molina, e-print gr-qc/0304053.

\bibitem{Pol} A. P. Polychronakos, e-print hep-th/0304135.

\bibitem{Bri2} A. M. van den Brink, e-print gr-qc/0304092.

\bibitem{LiHui} L. H. Xue, Z. X. Shen, B. Wang, and R. K. Su, e-print gr-qc/0304109.

\bibitem{CarKon} V. Cardoso, R. Konoplya, J. P. S. Lemos, e-print gr-qc/0305037.

\bibitem{BirCar} D. Birmingham, S. Carlip, and Y. Chen, e-print hep-th/0305113.

\bibitem{Swa} J. Swain, e-print gr-qc/0305073.

\bibitem{Pad} T. Padmanabham and A. Patel, e-print hep-th/0305165.

\bibitem{Bir} D. Birmingham, e-print hep-th/0306004.

\bibitem{Iiz} N. Iizuka, D. Kabat, G. Lifschytz, and D. A. Lowe, e-print hep-th/0306209.

\bibitem{Ber2} E. Berti, V. Cardoso, K. D. Kokkotas and H. Onozawa, e-print hep-th/0307013.

\bibitem{AndHow} N. Andersson and C. J. Howls, e-print gr-qc/0307020.

\bibitem{Hod1} S. Hod, Phys. Rev. Lett. {\bf 81}, 4293 (1998).

\bibitem{Nollert2} H. P. Nollert, Phys. Rev. D {\bf 47}, 5253 (1993).

\bibitem{Andersson} N. Andersson, Class. Quantum Grav. {\bf 10}, L61 (1993).

\bibitem{Muk} V. Mukhanov, JETP Lett. {\bf 44}, 63 (1986).

\bibitem{BekMuk} J. D. Bekenstein and V. F. Mukhanov, Phys. Lett. B
  {\bf 360}, 7 (1995).

\bibitem{Beken2} J. D. Bekenstein, Lett. Nuovo Cimento {\bf 11}, 467 (1974).

\bibitem{Beken3} J. D. Bekenstein in XVII Brazilian National Meeting
  on Particles and Fields, eds. A. J. da Silva et. al. (Brazilian
  Physical Society, Sao Paulo, 1996), J. D. Bekenstein in Proceedings of the VIII
  Marcel Grossmann Meeting on General Relativity, eds. T. Piran and
  R. Ruffini (World Scientific , Singapore, 1998).

\bibitem{Beken1} J. D. Bekenstein, Phys. Rev. D {\bf 7}, 2333 (1973).

\bibitem{Det} S. Detweiler, Astrophys. J. {\bf 239}, 292 (1980).

\bibitem{Ono} H. Onozawa, Phys. Rev. D {\bf 55}, 3593 (1997).

\bibitem{RegWheel} T. Regge and J. A. Wheeler, Phys. Rev. {\bf 108},
  1063 (1957).

\bibitem{Teukolsky} S. A. Teukolsky, Phys. Rev. Lett. {\bf 29}, 1114 (1972); 
Astrophys. J. {\bf 185}, 635 (1973).

\bibitem{Detwe} S. L. Detweiler, in Sources of Gravitational
  Radiation, edited by L. Smarr (Cambridge University Press,
  Cambridge, England, 1979).

\bibitem{Note1} To our best knowledge, the simple formulation of these coeficients in terms 
of the black-hole parametrs ($T_{BH}$ and $\Omega$) has not been done so far.

\bibitem{Note2} Teukolsky's angular equation belongs to the family of differential equations for 
angular prolate spheroidal wave functions. The behavior of the angular eigenvalues, for large $a \omega$, is 
well known \cite{Abr,Gia}, and is given by $A_{lm}=O(a\omega)$.

\bibitem{Abr} M. Abramowitz, {\it Handbook of Mathematical Functions}, (Dover 
Publications, inc., New York, 1965).

\bibitem{Gia} M. Giammatteo, e-print gr-qc/0303011.

\end{thebibliography}
\end{document}